\documentclass[preprints,article,accept,moreauthors,pdftex]{Definitions/mdpi}

\usepackage{caption}

\newcommand{\ctu}[1]{\hat{c}_{\mathbf{#1} \uparrow}}

\newcommand{\mctd}[1]{\hat{c}_{-\mathbf{#1} \downarrow}}

\newcommand{\cts}[1]{\hat{c}_{\mathbf{#1} \sigma}}

\newcommand{\ket}[1]{\left| #1 \right>}
\newcommand{\mv}[1]{\left< #1 \right>}

\newcommand{\mf}[1]{\mathbf{#1}}
\newcommand{\x}[1]{\xi _{\mf{#1}}}

\newcommand{\+}{\dagger}

\newcommand{\uk}[1]{u_{\mf{#1}}}
\newcommand{\vk}[1]{v_{{\mf{#1}}}}
\newcommand{\Dk}[1]{\Delta_{{\mf{#1}}}}
\newcommand{\Ek}[1]{E_{{\mf{#1}}}}

\newcommand{\rmf}{\textrm{F}}

\allowdisplaybreaks

\firstpage{1} 
\pubvolume{1}
\issuenum{1}
\articlenumber{0}
\pubyear{2022}
\copyrightyear{2022}
\externaleditor{Academic Editors: Sebastiano Pilati, Bruno Julia-Diaz, and Stefano Giorgini} 
\datereceived{} 
\dateaccepted{} 
\datepublished{} 
\hreflink{https://doi.org/} 

\Title{The $^1$S$_0$ Pairing Gap in Neutron Matter}

\TitleCitation{The $^1$S$_0$ Pairing Gap in Neutron Matter}

\Author{Stefano Gandolfi $^{1}$, Georgios Palkanoglou $^{2}$, Joseph Carlson $^{1}$, Alexandros Gezerlis $^{2}$, and Kevin~E.~Schmidt $^{3}$}

\AuthorNames{Stefano Gandolfi, Georgios Palkanoglou, Joseph Carlson, Alexandros Gezerlis, and Kevin~E.~Schmidt}

\AuthorCitation{Gandolfi, S.; Palkanoglou, G.; Carlson, J.; Gezerlis, A.; Schmidt K.E.}

\address{%
$^{1}$ \quad Theoretical Division, Los Alamos National Laboratory, Los Alamos, NM 87545, USA; stefano@lanl.gov (S.G.); carlson@lanl.gov (J.C.) \\
$^{2}$ \quad Department of Physics, University of Guelph, Guelph, ON N1G 2W1, Canada; gpalkano@uoguelph.ca (G.P.); gezerlis@uoguelph.ca (A.G.)\\
$^{3}$ \quad Department of Physics, Arizona State University, Tempe, AZ 85287, USA; Kevin.Schmidt@asu.edu}

\corres{Correspondence: stefano@lanl.gov}

\abstract{We report ab initio calculations of the $S$ wave pairing gap
in neutron matter calculated using realistic nuclear Hamiltonians that
include two- and three-body interactions. We use a trial state, properly optimized to capture the essential pairing correlations, from which we extract ground state properties by means of auxiliary field diffusion Monte Carlo simulations. We extrapolate our results to the thermodynamic limit by studying the finite-size effects in the symmetry-restored projected Bardeen-Cooper-Schrieffer (PBCS) theory and compare our results to other ab initio studies done in the past. Our quantum Monte Carlo results for the pairing gap show a modest suppression with respect to the mean-field BCS values. These results can be connected to cold atom experiments, via the unitarity regime where fermionic superfluidity assumes a unified description, and they are important in the prediction of thermal properties and the cooling of neutron stars.
}


\begin{document}
Strongly paired Fermi systems offer a unique regime for quantum many-body physics as their relevance spans many physical settings of various scales: From the structure of neutron stars (NSs) and the physics of neutron-rich nuclei to cold-atom experiments. Neutron matter (NM), one of the most strongly interacting Fermi systems found in nature, is an important ingredient of NSs, playing an essential role in their structure~\cite{Gandolfi:2015} while strongly interacting fermionic atoms are now routinely used in experiments shedding light on the properties of strongly interacting superfluids~\cite{Ketterle:2008}. While it initially appears different, the description of these systems can be unified via their proximity to the unitary Fermi gas, connecting atomic experiments on Earth to the NS matter.

Neutrons found in the inner crust of quiescent NSs are known to form $^1$S$_0$ pairs, turning low-density NM to a $S$ wave superfluid~\cite{Gandolfi:2015,Ramanan:2021,Dean:2003}. The correct description of such neutron fluids is integral to the understanding of NS physics. Properties of low-density NM can explain observations such as irregularities in the periods of NSs and their cooling~\cite{Page:2004, Page:2011, Yakovlev:2004, Watanabe:2017}, while the equation of state (EoS) of high-density NM impacts the mass-radius relations of NSs~\cite{Gandolfi:2015, Lattimer:2016} and the hydrodynamic description of their inner crust~\cite{Chamel:2017}. Neutron matter of the same densities is also found on the exterior of neutron-rich nuclei~\cite{Dean:2003,Gandolfi:2015}. Therefore, a correct description of low-density NM is crucial to our understanding of nuclear systems of various sizes.

Strongly interacting cold Fermi atoms have been the subject of many theoretical investigations~\cite{Pilati:2008,Strinati:2018,Forbes:2011,Galea:2016,Carlson:2011,Magierski:2009,Zielinski:2020, Tajima:2017}. Experimentally, they have been studied extensively since the beginning of the century, owing in part, to the simplicity of these experiments compared to those for their bosonic counterparts~\cite{Ketterle:2008}. In these cold atom experiments, the strength of the interaction can be tuned through Feshbach resonances to yield a specific scattering length. Many experimental studies of strongly interacting Fermi gases utilize a $^6$Li gas, which exhibits a very broad Feshbach resonance, with a vanishing effective range $r_\textrm{e}$~\cite{Bartenstein:2005}. This allows one to perform studies of atomic superfluids from close to the Bardeen-Cooper-Schrieffer (BCS) limit (small $-k_\rmf a$) to unitarity ($-k_\rmf a\gg 1)$, where $k_\rmf$ is the Fermi momentum and $a$ is the scattering length of the inter-particle interaction. For a comprehensive review of experimental techniques of cold Fermi atoms at unitarity, see Ref.~\cite{Ketterle:2008}.

The neutron-neutron (NN) interaction is in principle a very complicated one. At~large distances, it is described by the exchange of a pion, and at intermediate distances it is spin-dependent and attractive, dominated by a two-pion exchange, and it turns repulsive at short distances. However, at the very low densities found in the inner crust of NSs or the exterior of neutron-rich nuclei, the long inter-particle distance allows us to ignore most of these details, and capture the physics of the system with the scattering length and effective range of the NN interaction, while the repulsive core only guarantees that the system does not collapse to a higher-density state. This is known as the shape independence and it asserts that, at low energies, the two-body scattering phase-shift $\delta_0$ can be described solely by $a$ and $r_e$,
\begin{equation}
    \cot\delta_0(k)=-\frac{1}{a}+\frac{1}{2}r_ek^2+\cdots ~.
\end{equation}

The parameters $a$ and $r_{\rm e}$ can be thought of as the strength and the range of the interaction that drives the low-energy scattering.
Therefore, for low densities, we can model the NN interaction with any potential that reproduces the correct scattering length and effective range.

The unitarity limit refers to Fermi gases with an attractive interaction strong enough to create a bound state of vanishing bound energy. In terms of the scattering length and effective range, this corresponds to $-k_\rmf a = \infty$ and $k_\rmf r_{\rm e}=0$ making the inter-particle distance, introduced by the density, the only length scale of the system. Therefore, Fermi gases are expected to exhibit universal behavior at this limit~\cite{Ho:2004}. Past unitarity, for $k_\rmf a >1$, lies the Bose--Einsein condensation (BEC) regime, where pairing has grown strong enough to create bound molecular states. At a zero temperature, pairing is present throughout the three regimes, with unitarity being a smooth crossover between the BCS and BEC regime. For a discussion on the BCS-BEC crossover and how it can be used to connect cold atoms to nuclear systems, see Refs.~\cite{Strinati:2018, Ohashi:2020}.

\begin{figure}[H]
\includegraphics[width=0.95\columnwidth]{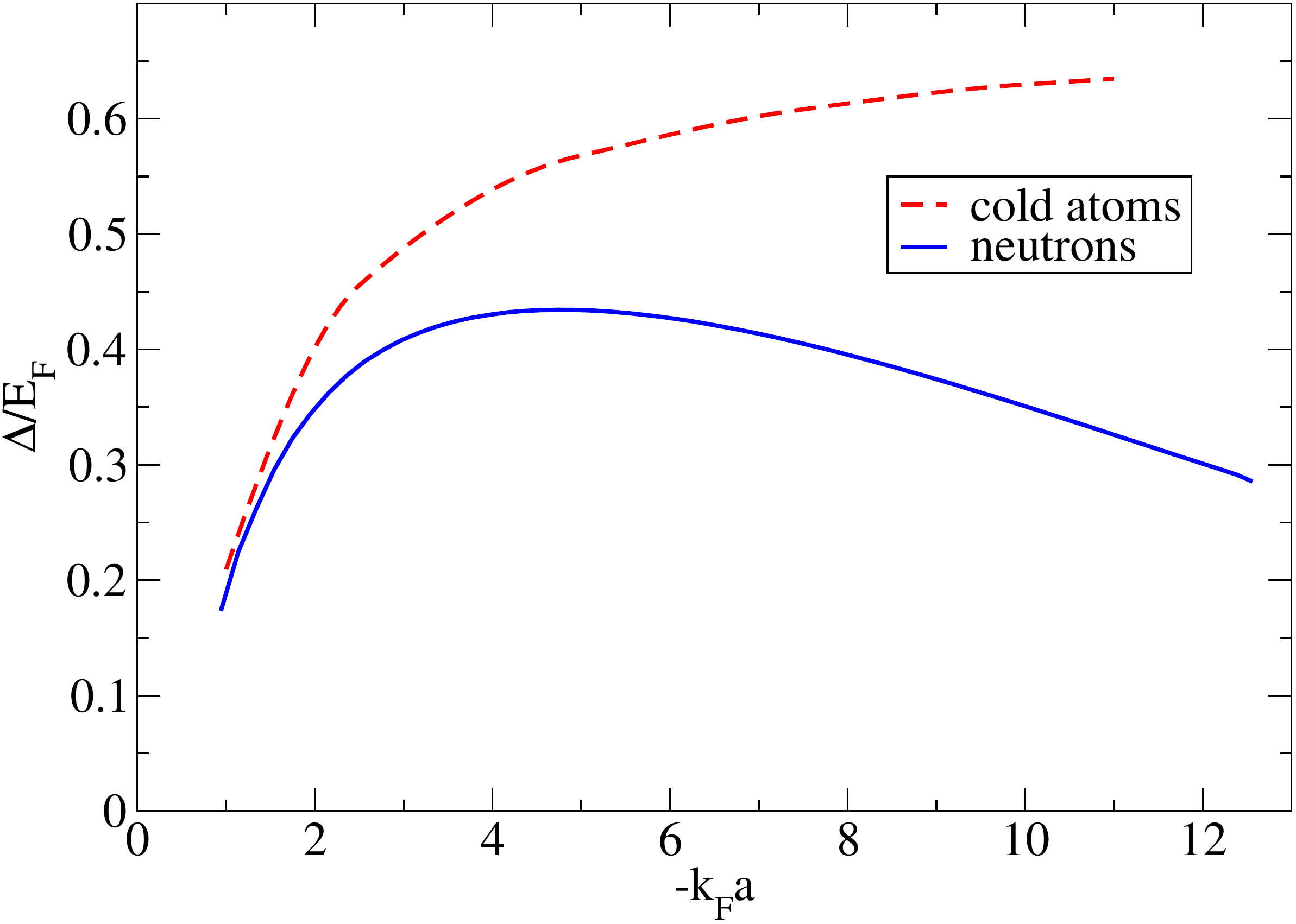}
\caption{The pairing gap in units of the Fermi energy for neutron matter (NM; blue solid line), and cold atoms (red dashed line), calculated solving the Bardeen-Cooper-Schrieffer (BCS) gap equations at the thermodynamic limit (TL) (see~\mbox{Section \ref{sec:bcs})}.} \label{fig:nm_ca_DEF}
\end{figure}

While the scattering length of cold atoms can be tuned arbitrarily close to unitarity (and beyond), NM comes with fixed $a$ and $r_{\rm e}$. In the densities considered here, the NN interaction exhibits a very long scattering length $a\approx-18.7~{\rm fm}$ being attractive enough to almost form bound states (dineutrons), while its effective range is finite, but much smaller than the scattering length, at $r_{\rm e}\approx2.7~{\rm fm}$. While the scattering length is much larger than the inter-particle spacing for low-density NM, very low densities are needed to turn the inter-particle spacing to be much larger than $r_{\rm e}$. Therefore, to the extent that the effects of a finite range can be neglected, neutron matter and cold atoms exhibit `universal' behavior; their properties depend only on the product of the Fermi momentum with the scattering length. This can be clearly seen in Figure~\ref{fig:nm_ca_DEF}, where for low densities (and large inter-particle distances) NM and cold-atoms produce identical pairing gaps. This changes as the density increases and the effects of the finite range NN interaction start becoming important. 

The exponential suppression of pairing on the BCS limit allows for a mean-field description of superfluidity, namely the BCS theory (see Section~\ref{sec:bcs}). While the mean-field approach can be applied to stronger pairing as well, it is expected to be only qualitatively correct. For an accurate description of strongly paired systems, such as the unitary gas, we have to look past the mean-field approach. One way is paved by the various beyond-mean field theories, which include correlations neglected by the mean-field treatment (e.g., see Refs.~\cite{Ding:2016, Pavlou:2017,Krotscheck:1980,Fan:2018,Cao:2006,Wambach:1993}). Another path are ab initio approaches where the problem is tackled from first principles without any uncontrolled approximations. We have performed auxiliary field diffusion Monte Carlo (AFDMC) calculations of the ground-state properties of superfluid NM for an improved variational wavefunction. We compare our results to previous exact diffusion Monte Carlo (DMC) calculations for very low-density NM and we extend our results to higher densities. Finally, we compare our results to these predicted by the BCS theory and its particle-conserving version, for the same densities. The rest of the manuscript is organized as follows: In Section~\ref{sec:bcs}, we present a brief overview of the BCS theory and related techniques which provide a qualitative understanding of pairing, in Section~\ref{sec:qmc} we present the overview of the AFDMC approach that was used and a comparison with the DMC in past calculations, and we conclude with Sections~\ref{sec:results-energy} and \ref{sec:results-gap},  presenting and discussing our results for the pairing gap and the EoS of low-density~NM.

\section{Superfluid NM and the BCS Theory}\label{sec:bcs}
\subsection{Overview of BCS Theory}
The BCS theory describes superfluid fermionic gases as condensates of Cooper pairs. The ground state of these systems is the fully paired condensate that minimizes the free energy corresponding to the pairing Hamiltonian, i.e.,~the one neglecting all normal state~interactions:
\begin{equation}
    \hat{H}_{\textrm{pair}} = \sum_{\mf{k}\sigma}\epsilon_\mf{k}\cts{k}^\+ \cts{k} + \sum_{\mf{k}\mf{l}} V_{\mf{k}\mf{l}}\ctu{k}^\+\mctd{k}^\+\mctd{l}\ctu{l}~. \label{eq:H_pair}
\end{equation}

Here, $\epsilon_{\mf{k}}$ signifies the single-particle energy of a free particle with momentum $\mf{k}$, in a box of length $L$ under Periodic Boundary Conditions (PBC), and the matrix elements $V_{\mf{k}\mf{l}}$ are those of the pairing interaction, the attractive interaction responsible for the creation of the Cooper pairs. The BCS ground state for systems with even and odd particle numbers, can be written as:
\begin{align}
    \ket{\psi_{\textrm{BCS}}}_{\rm even} &= \prod_{\mf{k}}\left(\uk{k} + \vk{k}\ctu{k}^\+\mctd{k}^\+\right)\ket{0}~, \label{eq:BCS_gs} \\
    \ket{\psi_{\textrm{BCS}}^{\mf{b}\gamma}}_{\textrm{odd}} &= \hat{c}_{\mf{b}\gamma}^\+\prod_{\mf{k}\neq \mf{b}}\left(\uk{k} + \vk{k}\ctu{k}^\+\mctd{k}^\+\right)\ket{0}~. \label{eq:BCS_gs_o}
\end{align}
where, the distribution $\vk{k}$ ($\uk{k}$) is the probability amplitude for (not) finding a pair with momenta and spin $\mathbf{k} \uparrow$, $-\mf{k}\downarrow$ and so it is normalized to unity, $\uk{k}^2+\vk{k}^2 = 1$. Hence, the distribution $v_\mf{k}$ completely defines the BCS ground state and it is determined so that the state in Equation~(\ref{eq:BCS_gs}) minimizes the free energy. This condition leads to the BCS gap equations which determine the gap function $\Delta_\mf{k}$, corresponding to half the binding energy of a $\mf{k}$-pair, and the quasiparticle excitation energy, 
\begin{equation}
    \Ek{k} = \sqrt{\left(\epsilon_{\mf{k}}-\mu\right)^2 + \Dk{k}^2}~, \label{eq:Ek}
\end{equation}
where $\mu$ is the chemical potential. These in turn define the distribution $\vk{k}=[1-(\epsilon_\mf{k}-\mu)/{\Ek{k}}]/2$.

A key signature of pairing correlations is the existence of a minimum non-zero energy required to create an excitation by breaking apart a pair. This is called the pairing gap and, in the BCS theory, it corresponds to the minimum of the quasi-particle excitation energy:
\begin{align}
    \Delta_{\textrm{MF}}=\textrm{min}_\mf{k}\Ek{k}~. \label{eq:gap_mf}
\end{align}

Here the subscript ``MF'' refers to the mean-field nature of the BCS theory. This can be seen by performing a Bogolyubov transformation and describing the theory in terms of non-interacting quasi-particles moving in an average field ~\cite{Rickayzen:book}.

\subsection{BCS Theory for NM and Cold Atoms}

The BCS theory can provide a phenomelogical description of the $^1$S$_0$ pairing in low-density NM, where neutrons form spin-singlet ($S=0$) pairs. In such a description, we can model the neutron-neutron (NN) interaction with a simple potential that is tuned to reproduce the $^1$S$_0$ scattering length ($a\approx-18.5~\rm{fm}$) and effective range ($r_{\rm{e}}\approx 2.7~\rm{fm}$) of the NN interaction: Such a potential should produce indistinguishable results for low-density NM where the details of the functional form of the interaction are irrelevant and the physics is captured by $a$ and $r_{\rm e}$. We choose the form of a modified P\"{o}schl--Teller potential:
\begin{equation}
    V(r) = -\frac{\hbar^2}{m}\frac{\lambda(\lambda-1)\beta ^2}{\cosh^2(\beta r)}~, \label{eq:potential}
\end{equation}
where the parameters $\lambda$ and $\beta$ are tuned to reproduce the $^1$S$_0$ scattering length and effective range of NM. This potential has been used successfully before in phenomenological studies of pairing in NM~\cite{Palkanoglou:2020, Palkanoglou:2021, Gezerlis:2010}.

For cold atoms, the interaction's scattering length can be tuned using Feshbach resonances~\cite{Ketterle:2008}. For low densities, the functional form of the interaction is irrelevant and the inter-atomic potential can be modeled by Equation~(\ref{eq:potential}). To study atoms close to unitarity, we tune the parameters $\beta$ and $\lambda$ to yield a small effective range (smaller than the inter-particle distance) and the appropriate scattering length. Therefore we can treat NM and cold atoms, at low densities, on equal footing when formulating a description within the BCS theory.

Since we aim for a description of the $^1$S$_0$ pairing, all BCS equations need to be expanded in partial waves where only the $S$ wave is to be kept, leading to the angle-averaged BCS gap equations, for even systems,
\begin{align}
    \Delta_{\rm{even}}(k) &= -\frac{2\pi}{L^3}\sum_p M(p)V_0(k,p)\frac{\Delta (p)}{E(p)}~, \label{eq:gap1_A} \\
    \mv{\hat{N}}&=\sum_pM(p)\left(1-\frac{\epsilon(p)-\mu}{E(p)}\right)~, \label{eq:gap2_A}
\end{align}
and for odd systems,
\begin{align}
    \Delta_{\rm{odd}}(k) &= -\frac{2\pi}{L^3}\sum_{p\neq b} M(p)V_0(k,p)\frac{\Delta (p)}{E(p)}~, \label{eq:gap1_o_A}\\
    \mv{\hat{N}}-1&=\sum_{p\neq b}M(p)\left(1-\frac{\epsilon(p)-\mu}{E(p)}\right)~, \label{eq:gap2_o_A}
\end{align}
and the angle-averaged version of the ground state energies,
\begin{align}
    E^{\rm{BCS}}_{\rm{even}}(N) &= \sum _p M(p) \epsilon(p) 2v(p)^2 + \notag\\
    &\quad +\frac{4\pi}{L^3}\sum _{kp}M(k)M(p) V_0(k,p)u(k)v(k)u(p)v(p)~, \label{eq:energy_A}\\
    E^{\rm{BCS}}_{\rm{odd}}(N) &= \sum _{p\neq b} M(p) \epsilon(p) 2v(p)^2 + \epsilon(b) \notag\\
    &\quad +\frac{4\pi}{L^3}\sum _{kp\neq b}M(k)M(p) V_0(k,p)u(k)v(k)u(p)v(p)~. \label{eq:energy_o_A}
\end{align}

In this angle-averaged expression of BCS, the population function $M(k)$ counts the number of $\mf{k}$ states in the momentum shell with $|\mf{k}|=k$, and all functions of $\mf{k}$ have been replaced by their angle-averaged counterparts which are functions of just the momentum's magnitude $k$. It should be noted that, in the expressions for odd systems, only a single state with momentum $\mf{b}$ is excluded from the sum rather than the entire $|\mf{k}|=b$ shell. The potential $V_0$ is that of the $^1$S$_0$ channel of the potential in Equation~(\ref{eq:potential}), i.e.,~the $S$ wave term in its partial wave expansion, $V_0(k,p) = \int_0^\infty drr^2j_0(kr)V(r)j_0(pr)$, where $j_0$ is the zero-th order spherical Bessel function. For a complete derivation of the angle-averaged version of BCS, see Ref.~\cite{Palkanoglou:2020}.

While the formulation of BCS presented until this point describes a finite number of neutrons or atoms in a cubic box of length $L$, a formulation for a superfluid system at the Thermodynamic Limit (TL), such as NM found in the inner crust of NSs, can be retrieved by taking the limit of $L\to \infty$ while keeping the particle density constant, i.e.,~$n = \left<N\right>/L^3=\textrm{const}$. The distinction between even and odd systems is irrelevant at the TL therefore \mbox{Equations~(\ref{eq:gap1_A}),~(\ref{eq:gap2_A}), and~(\ref{eq:energy_A})}, and \mbox{Equations~(\ref{eq:gap1_o_A}),~(\ref{eq:gap2_o_A}),  and~(\ref{eq:energy_o_A})} have the same large-$N$ limit.

\subsection{PBCS: Particle-Number Projected BCS}

The BCS ground state in Equations~(\ref{eq:BCS_gs})~and~(\ref{eq:BCS_gs_o}) does not conserve the particle number and, therefore, it describes a linear combination of states with an average particle number of $\mv{N}$. Quantum Monte Carlo  calculations deal with states of a well-defined particle number. To connect such calculations to values at the TL, we need to study the superlfuid FSE in a particle-conserving theory. Since, particle-number conservation is a symmetry of the Hamiltonian (cf. Equation~(\ref{eq:H_pair})), we can use symmetry-restoration techniques to restore this symmetry in the BCS ground state, which amounts to projecting out of the ground state the component that corresponds to the proper eigenstate of the number operator~\cite{Dietrich:1964, Bayman:1960}:
\begin{align}
    \ket{\psi_N}&=C\int _0^{2\pi} d\phi e^{-i\frac{N}{2}\phi} \prod_{\mf{k}}\left(\uk{k}+e^{i\phi}\vk{k}\ctu{k}^\+\mctd{k}^\+\right)\ket{0}~, \label{eq:pgs}\\
    \ket{\psi^{\mf{b}\gamma}_N}&=C(\mf{b})\hat{c}^\+_{\mf{b}\gamma}\int _0^{2\pi}d\phi e^{-i\frac{N}{2}\phi} \prod_{\mf{k}}\left(\uk{k}+e^{i\phi}\vk{k}\ctu{k}^\+\mctd{k}^\+\right)\ket{0}~. \label{eq:pgs_o}
\end{align}

Here, $C$ and $C(\mf{b})$ are the normalization of the projected BCS state for even and odd systems, respectively. In principle, one must treat the states in Equations~(\ref{eq:pgs})~and~(\ref{eq:pgs_o}) as variational wavefunctions and determine the distributions $\uk{k}$ and $\vk{k}$ that minimize the energy of each state, an approach called variation after projection (VAP) or full BCS (FBCS). An alternative route is to use the distributions $\uk{k}$ and $\vk{k}$ that solve the BCS gap equations, which amounts to the projection after variation (PAV) or projected BCS (PBCS). The PBCS ground states are an approximation of the FBCS ones, in the sense that the distributions $\uk{k}$ and $\vk{k}$ are not optimized to yield minimum energy. However, the error introduced is small for strongly paired systems~\cite{Bayman:1960}, such as NM.
Dealing with $S$ wave superfluidity, we must again isolate the $S$ wave terms from a partial wave expansion, and so the energy of the PBCS ground states for even and odd systems are:
\begin{align}
    E_{\rm{even}}^{\rm{PBCS}}(N) &= \sum _k M(k)\epsilon(k) 2v(k)^2\frac{R_1^1(k)}{R_0^0} \notag \\  &\quad +\sum_{kp}M(k)M(p)V_0(k,p)u(k)v(k)u(p)v(p)\frac{R_1^2(k\,p)}{R_0^0}~, \label{eq:energy_pbcs.s}\\
    E_{\rm{odd}}^{\rm{PBCS}}(b;N) &= \sum _{k\neq b} M(k)\epsilon(k)2v^2(k)\frac{R_1^2(b\, k)}{R^1_0(b)} \notag \\  & \quad +\sum_{kp}M(k)M(p)V_0(k,p)u(k)v(k)u(p)v(p)\frac{R_1^3(b\,k\,p)}{R^1_0(b)}~. \label{eq:energy_pbcs_o.s}
\end{align}
where we defined the residuum integrals,
\begin{align}
    R^m_n(\mf{k}_1\,\mf{k}_2\dots\mf{k}_m) = \int_0^{2\pi} \frac{d\phi}{2\pi}e^{-i(\frac{N}{2}-n)\phi} \prod_{\mf{k}\neq \mf{k}_1,\mf{k}_1,\dots \mf{k}_m}\left(\uk{k}^2 + e^{i\phi}\vk{k}^2\right).
\end{align}

\begin{figure}[H]%
\widefigure
\begin{minipage}{0.45\columnwidth}%
{\includegraphics[width=1\columnwidth,clip=]{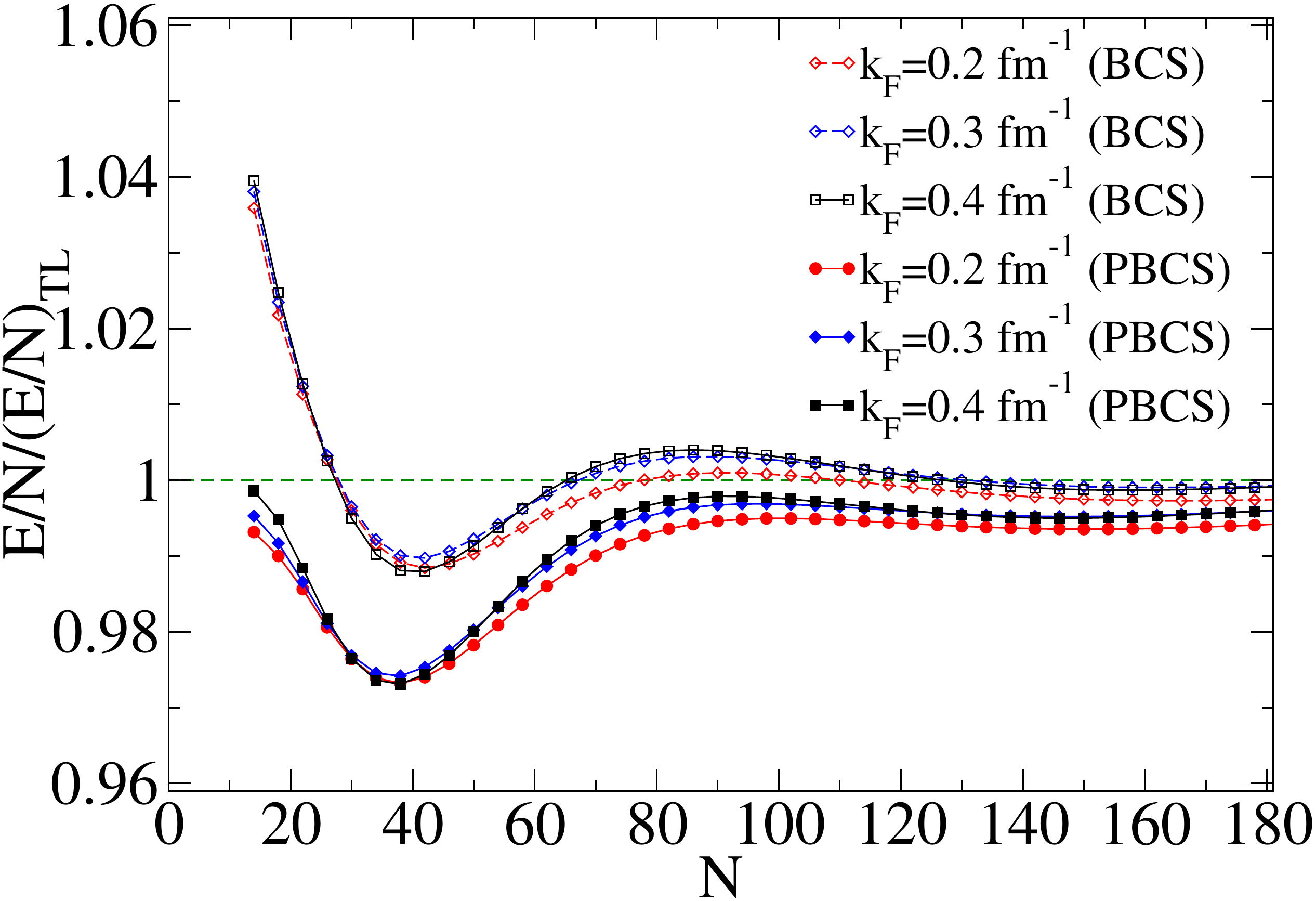}}%
\end{minipage}
\qquad
\begin{minipage}{0.45\columnwidth}%
\includegraphics[width=1\columnwidth,clip=]{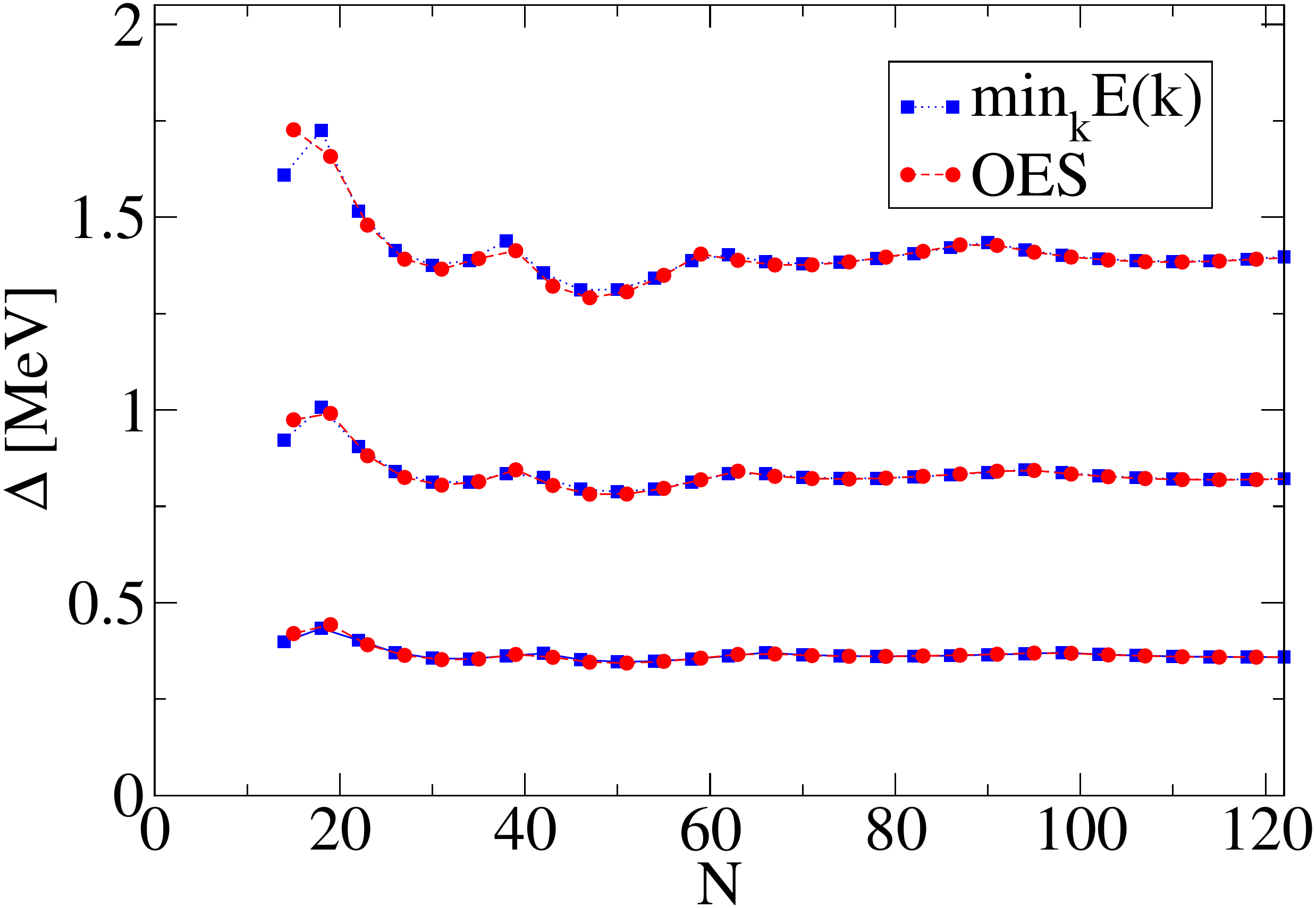}
\end{minipage}%
\caption{{\textbf{(Left~panel)}:} The finite-size effects (FSE) for the energy in BCS and Projected BCS (PBCS) for densities corresponding to $k_\textrm{F} = 0.2$, $0.3$, $0.4~\textrm{fm}^{-1}$. (\textbf{Right panel}): The FSE of the pairing gap in BCS as the minimum of the quasiparticle excitation energy (blue squares) and in PBCS as the odd-even staggering (OES), corresponding to $k_\textrm{F} = 0.2$, $0.3$, $0.4~\textrm{fm}^{-1}$, in ascending order.}
\label{fig:fse}
\end{figure}

The energy per particle calculated with Equations~(\ref{eq:energy_pbcs.s}) and~(\ref{eq:energy_pbcs_o.s}), or Equations~(\ref{eq:energy_A}) \mbox{and~{(\ref{eq:energy_o_A})}}, for a given density is not equal to the TL value of the energy, as is the case with any intensive quantity of a finite system. This is known as the finite-size effects (FSE) and they can be seen in the left panel of Figure~\ref{fig:fse}. By studying the FSE of intensive quantities, we can prescribe extrapolation schemes to the TL and estimate their accuracy. In the case of superfluidity, pairing tends to create smooth FSE for the energy with no abrupt changes, compared to the FSE of the free Fermi gas. This can be most clearly seen in a comparison of the superfluid kinetic energy with that of the free Fermi gas, plotted in Figure~\ref{fig:kinetic}. This can be attributed to the pairing's smearing of the Fermi surface (see Figure~\ref{fig:v2k}). When studying the FSE, the particle-number projection of PBCS should be seen as separating the contribution of a system with $N$ neutrons from the linear combination of systems with average particle-number $\left< N\right>$, that is the BCS state.  As such, the PBCS curves in \mbox{Figure~\ref{fig:fse} and \ref{fig:kinetic}} represent more well-defined FSE curves compared to the BCS ones. A detailed study of the FSE in BCS and PBCS for NM was carried out in Ref.~\cite{Palkanoglou:2020}. 
The prescription of the pairing gap defined in the BCS theory in Equation~(\ref{eq:gap_mf}) cannot be applied in the PBCS theory. Alternatively, one can define the odd-even staggering (OES), 
\begin{equation}
    \Delta (N) = \frac{(-1)^N}{2}\left[2E(N) - E(N+1)-E(N-1)\right]~, \label{eq:oes}
\end{equation}
inspired by the odd-even mass staggering of nuclei. It has been demonstrated that for the $^1$S$_0$ pairing gap in NM, $\Delta _{\rm{MF}}$ and $\Delta$ are probing the same physical quantity and \mbox{Equations~\mbox{(\ref{eq:gap_mf})~and~(\ref{eq:oes})}} can be used interchangeably~\cite{Palkanoglou:2020}, as shown in the right panel of \mbox{Figure~\ref{fig:fse}}. Since the pairing gap is an intensive quantity, it generally suffers from larger FSE than the energy (see the right panel of Figure~\ref{fig:fse}).

\clearpage
\end{paracol}
\begin{paracol}{2}
\switchcolumn

\begin{figure}[H]

   \includegraphics[width=.95\linewidth]{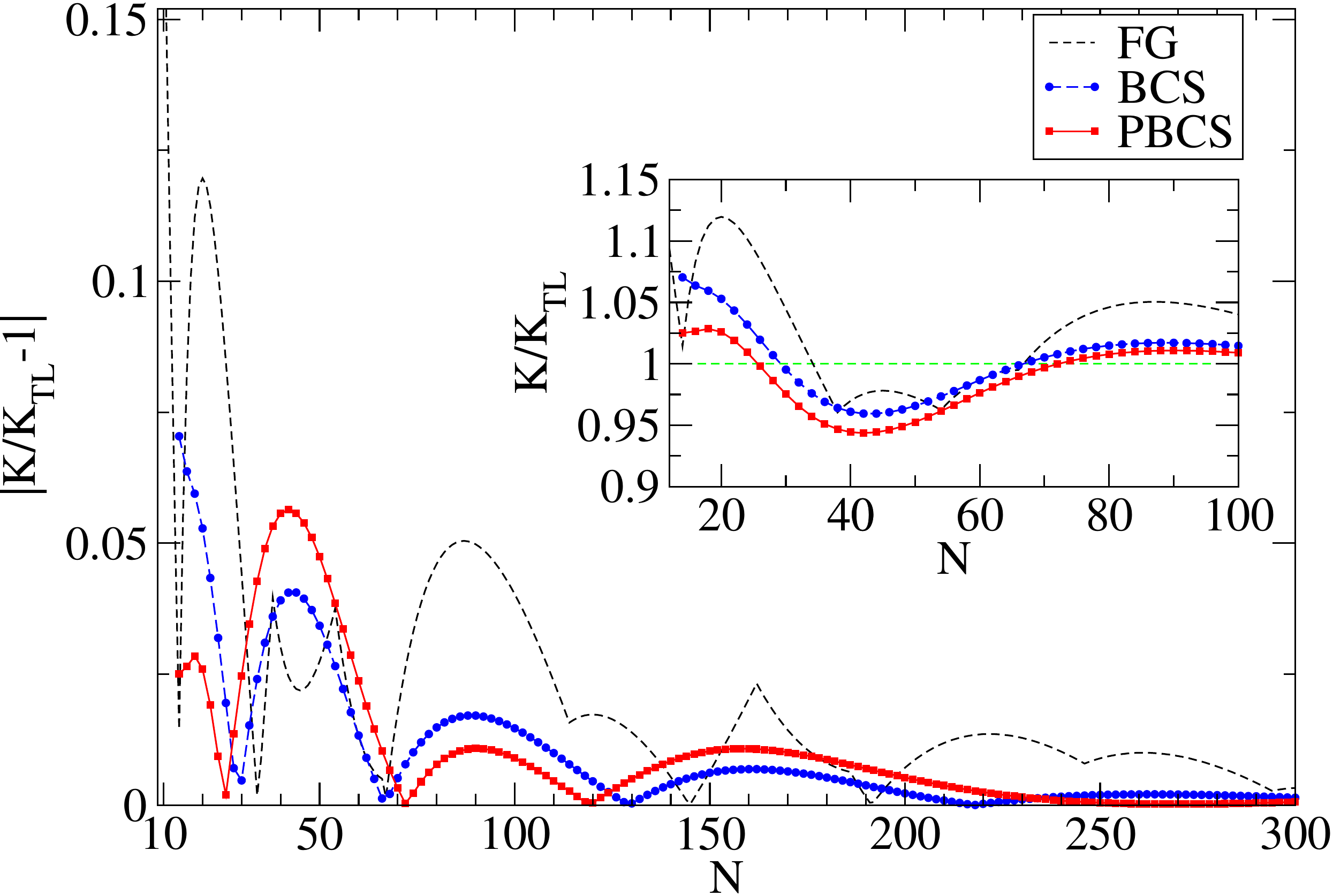}
\caption{The FSE of the superfluid kinetic energy compared to those of the free Fermi gas. The inset focuses on the region of $N=47$, the region explored by the auxiliary field diffusion Monte Carlo (AFDMC) calculations in Section~\ref{sec:results-energy}.}
\label{fig:kinetic}
\end{figure}

\begin{figure}[H]
 
  \includegraphics[width=.95\columnwidth]{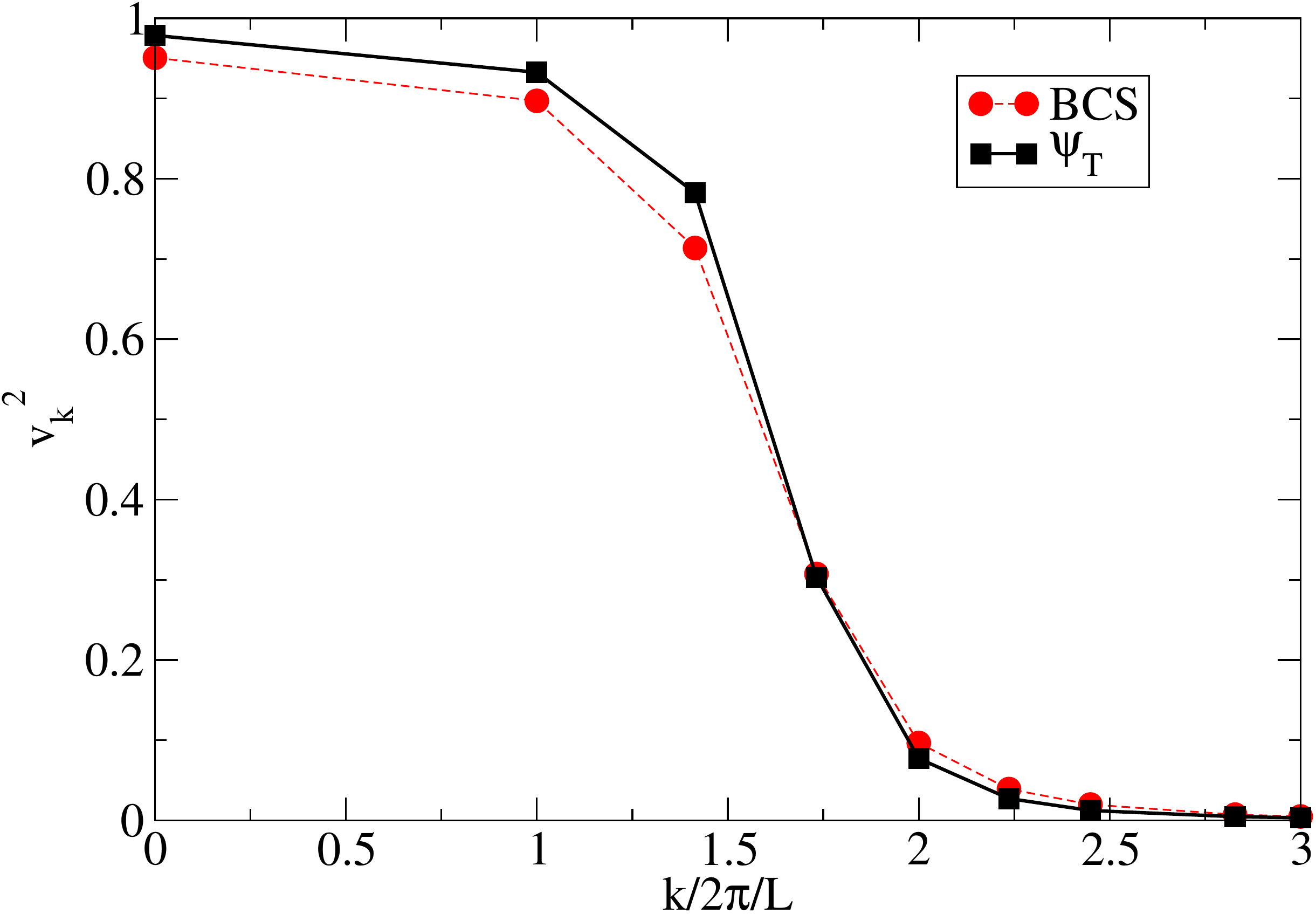}
\caption{The pair probability distribution used in this work (solid line) and the pairing function of the BCS ground state (dashed line) for $N=40$ particles at $k_\textrm{F}=0.4~\textrm{fm}^{-1}$.} \label{fig:v2k}
\end{figure}

\section{Ab Initio: DMC and AFDMC}\label{sec:qmc}

While the mean-field description, given by BCS, provides a qualitative understanding of strongly paired systems, accuracy demands that we treat pairing correlations from first principles. For NM, this can be done by numerically solving Schr{\"o}dinger's equation for the nuclear Hamiltonian, to find the ground state of a finite number of neutrons, and then extrapolating the results to the TL. We employ the non-relativistic nuclear Hamiltonian:
\begin{equation}
\label{hamiltonian}
H=-\frac{\hbar^2}{2m}\sum_{i=1}^N\nabla_i^2+\sum_{i<j}v_{ij}
+\sum_{i<j<k}V_{ijk} \,,
\end{equation}
where $m$ is the mass of the neutron, and $v_{ij}$ and $V_{ijk}$ are two-
and three-body potentials. All the results presented in this paper have been 
obtained using the Argonne AV8' and the Urbana-IX (UIX)~\cite{Wiringa:2002,Pudliner:1995}. The AV8' belongs to the Argonne family of realistic two-nucleon potentials, which are generated by high-precision fitting of experimental scattering data. The functional form of the AV8',
\begin{equation}
    v_{ij}=\sum_{p=1}^{8}v_p(r_{ij})O^{(p)}(i,j)~\label{eq:av8}~,
\end{equation}
contains eight two-particle operators: The four central components $\boldsymbol{1},\, \boldsymbol{\tau}_i\cdot\boldsymbol{\tau}_j,\,\boldsymbol{\sigma}_i\cdot\boldsymbol{\sigma}_j,\,(\boldsymbol{\tau}_i\cdot\boldsymbol{\tau}_j)(\boldsymbol{\sigma}_i\cdot\boldsymbol{\sigma}_j)$, the tensor $S_{ij}$ and the tensor-$\tau$ $S_{ij}(\boldsymbol{\tau}_i\cdot \boldsymbol{\tau}_j)$ components, the spin-orbit $\boldsymbol{L}_{ij}\cdot\boldsymbol{S}_{ij}$, and the spin-orbit-$\tau$ $\boldsymbol{L}_{ij}\cdot\boldsymbol{S}_{ij}(\boldsymbol{\tau}_i\cdot \boldsymbol{\tau}_j)$ components (where $S_{ij}=3(\boldsymbol{\sigma}_i\cdot\hat{r}_{ij})(\boldsymbol{\sigma}_j\cdot\hat{r}_{ij})-(\boldsymbol{\sigma}_i\cdot\boldsymbol{\sigma}_j)$ and $\boldsymbol{L}_{ij}$, $\boldsymbol{S}_{ij}$ are the relative angular momentum and the total spin of the particle $ij$). The UIX is a three-body potential,
\begin{equation}
    V_{ijk} = V_{2\pi}+V_{\rm R}~, \label{eq:uix}
\end{equation}
which describes the exchange of two pions between three nucleons
via a spin-isospin dependent term~\cite{Fujita:1957}
and it is fit to reproduce the correct triton energy in Green's function Monte Carlo calculations and the expected saturation energy of nuclear matter in the Fermi hypernetted-chain approximation~\cite{Pudliner:1995}. The remaining term is a phenomenological part that sums other neglected terms. We have also considered two- and three-body local interactions 
constructed within chiral effective field theory~\cite{Gezerlis:2014,Lynn:2016}, 
but the results are very similar.
This is expected as this study is dedicated to low-density NM. The ground-state is calculated using the AFDMC method, and then compared to earlier ab initio results obtained by the DMC method. Both of these methods are members of the QMC family of ab initio approaches, which solve the many-body Schr{\"o}dinger's equation stochastically. 

The DMC approach is a projector method, which uses imaginary-time propagation to extract the ground-state of a Hamiltonian from a trial state. The method relies on the fact that, in imaginary time, Schr{\"o}dinger's equation turns into a diffusion equation where the large-time limit of any initial state (not orthogonal to the ground state) is the ground state of the system:
\begin{align}
    \lim_{\tau \to \infty}e^{-(\hat{H}-E_{\rm T})\tau}\ket{\Psi _T}\propto \ket{\Phi _0}.
\end{align}

The speed of the convergence depends on the choice of the trial state, which should capture the qualitative features of the problem at hand. In practice, the DMC method is applied by distributing particles according to the trial wavefunction and then propagating them in space by sampling the short-time propagator. The spin of each particle is considered `frozen', i.e.,~different spin-projections define different species of particles. This means that this method is most suitable for systems with spin-independent interactions. The antisymmetry of the fermionic wavefunctions creates regions where the trial wavefunction is negative, which complicates the interpretation of Schr{\"o}dinger's equation as a diffusion equation. This is known as the fermion sign problem and in DMC, it is typically addressed by fixing the nodal surface of the wavefunction ($\Psi=0$) to be the same as that of the trial wavefunction. This corresponds to separating the simulation space in regions where the trial wavefunction has a definite sign (nodal pockets) and evolving each one independently. Therefore, the large-time limit of the initial configuration of particles corresponds to the state with the lowest energy and the same nodal surface as the trial wavefunction. The energy of this state provides an upper-bound to the true ground state energy. The DMC method has been used in the past to calculate the $^1$S$_0$ pairing gap of low-density NM and cold atoms~\cite{Gezerlis:2008}. For a review of DMC, see Ref.~\cite{Foulkes:2001}.

The AFDMC approach is built on the same underlying principle as the DMC method: A ground state is projected out of a trial wavefunction via imaginary-time propagation. In contrast to DMC, the AFDMC time propagation can alter the spin-projection of the particles. When done naively, the inclusion of spin degrees of freedom has an unfavorable scaling behavior. This is mediated in AFDMC by applying a Hubbard--Stratonovich transformation to the short-time propagator, expressing the action of an operator $\exp{(-\lambda \hat{O}^2/2})$ as: 
\begin{equation}
    e^{-\lambda \hat{O}^2/2} = \frac{1}{\sqrt{2\pi}}\int_{-\infty}^\infty dx e^{-{x^2}/{2}}e^{x\sqrt{-\lambda}\hat{O}}~,
\end{equation}
thus improving the scaling behavior at the cost of additional integrations over auxiliary fields. 
This makes AFDMC the method of choice for systems with spin-dependent interactions, like nuclear systems (for a more detailed description of AFDMC, see Refs.~\cite{Carlson:2015,Lonardoni:2018}). This method has been applied in the past to 
calculate the energy and pairing gap of low density neutron 
matter~\cite{Fabrocini:2005,Gandolfi:2008,Gandolfi:2009}.

The trial wavefunction $\Psi _T$ of $N$ neutrons in AFDMC is typically of the form:
\begin{equation}
\label{psi_T}
\Psi_T(R,S)=\left[\prod_{i<j}f(r_{ij})\right] \Phi(R,S) \,,
\end{equation}
where $R$ and $S$ represent the $3N$ spatial coordinates and $3N$ up- and down-spin
components of the neutrons, and $f(r)$ is a two-body spin-independent correlation (see Ref.~\cite{Carlson:2015} for details).
The antisymmetric part $\Phi$ of the trial wave function is usually given
by the ground state of non-interacting Fermions (Fermi gas), which is
written as a Slater determinant of single particle functions. 
In the case of superfluid systems, the function $\Phi$ must include pairing correlations.
For spin-independent interactions, like ultra-cold Fermi gases, pairing correlations
can be included by using a Slater determinant~\cite{Carlson:2003}. However, in the case of a 
realistic nuclear interaction that can exchange the spin of neutrons, a pfaffian 
wave function must be used instead:
\begin{equation}
{\rm Pf}A={\cal A}[\phi(1,2),\phi(3,4) \dots \phi(N-1,N)]~.\label{eq:pfaffian}
\end{equation}

The details on how to calculate the pfaffian above are given in Ref.~\cite{Gandolfi:2009}.

The pairing functions $\phi(r)$ are:
\begin{align}
\phi(i,j)&= \sum_\alpha c_\alpha \exp\left[i \mathbf{k}_\alpha\cdot\mathbf{r_{ij}}
\right] \chi(s_i,s_j)~,
\label{eq:phi}
\end{align}
where sum over $\alpha$ indicates the $k$-space shells of the cube with $\mf{k}$ values:
\begin{equation}
\mf{k}= \frac{2\pi}{L}
 (n_x \hat x + n_y \hat y + n_z \hat z)
\end{equation}
for integer $n_x$, $n_y$, and $n_z$, and $L$ is the simulation box size.
The function $\chi$ is the spin-singlet wave function for two neutrons:
\begin{equation}
\chi(s_i,s_j)= \frac{1}{\sqrt{2}}
\left(\langle s_i s_j |\uparrow \downarrow\rangle
-\langle s_i s_j |\downarrow \uparrow\rangle\right) \,.
\end{equation}

Note that if the pairing coefficients
$c_\alpha$ are zero for all $|\mf{k}_\alpha| > k_F$, the pfaffian in Equation~(\ref{eq:pfaffian}) turns into a Slater
determinant of spin-up and spin-down neutrons filling the Fermi sea, and
the pfaffian form goes over to the normal liquid state.

We calculate the superfluid pairing gap and the EoS for the ground state of NM. The pairing gap is evaluated by taking the difference of the total energy of systems with even and odd particle numbers, as per Equation~(\ref{eq:oes}), which makes the calculation very sensitive to the error-bars of the energy. The AFDMC method involves the constrained-path approximation to control the sign problem, and the results depend upon the quality of the trial wave function used to model the system.
In several cases, an unconstrained path evolution is possible, giving exact results
for the ground-state energy within (usually large) statistical error bars~\cite{Lonardoni:2018}.
Thus, unconstrained-path calculations of the pairing gap 
would give very large error bars.
In addition, it is reasonable to assume that the quality of the constrained-path approximation
is similar for systems with $N-1$, $N$, and $N+1$ neutrons, so the systematic uncertainties
would cancel. However, the results for the energy, and in particular for the pairing gap,
ultimately depend on the trial wave function.

In the old calculations of Refs.~\cite{Fabrocini:2005,Gandolfi:2008,Gandolfi:2009},
the parameters $c_\alpha$ were
chosen by performing a correlated basis function 
calculation~\cite{Fabrocini:2005,Fabrocini:2008}. Old results showed a pairing gap very close to the one predicted by the BCS theory, and
predicted a higher EoS with respect to other~calculations.

In this study, we improved the trial wave function by performing
a variational search of the optimal $c_\alpha$ parameters using the
stochastic reconfiguration method~\cite{Sorella:2001}. As will be discussed in Sections~\ref{sec:results-energy} and \ref{sec:results-gap}, this yields a state that captures the appropriate correlations of superfluid NM and thus yields a ground state with lower energy and results consistent with the accurate DMC calculations at
very low-densities performed by Gezerlis and Carlson~\cite{Gezerlis:2008,Gezerlis:2010}. The pairing functions in Equation~(\ref{eq:phi}), once properly normalized, can be connected to a BCS state, like the one in Equation~(\ref{eq:BCS_gs}), through the $c_\alpha$ parameters,
\begin{equation}
    c_\alpha =\frac{v_{\mf{k}_\alpha}}{u_{\mf{k}_\alpha}}~.\label{eq:ctov}
\end{equation}

This allows us to compare the pairing functions and pair probability distributions corresponding to the pairing function of Equation~(\ref{eq:phi}) for $40$ neutrons to those of the BCS ground state in Figure~\ref{fig:v2k} and \ref{fig:pair-func}. We find that the BCS state defined by the $c_\alpha$ parameters yields a pair probability distribution with a slightly higher spread around the Fermi surface, describing higher pairing correlations. This is reflected in the optimized state's pair function which decreases slowly with the particle separation (see Figure~\ref{fig:pair-func}). This suggests that a pair of particles remains correlated at larger separations. Additionally, we find that the optimized state's pair function is less isotropic compared to the BCS ground state, owing to the non-central components of the nuclear force described by the richer operator-structure of the AV8' potential.

\begin{figure}[H]
\includegraphics[width=.95\columnwidth]{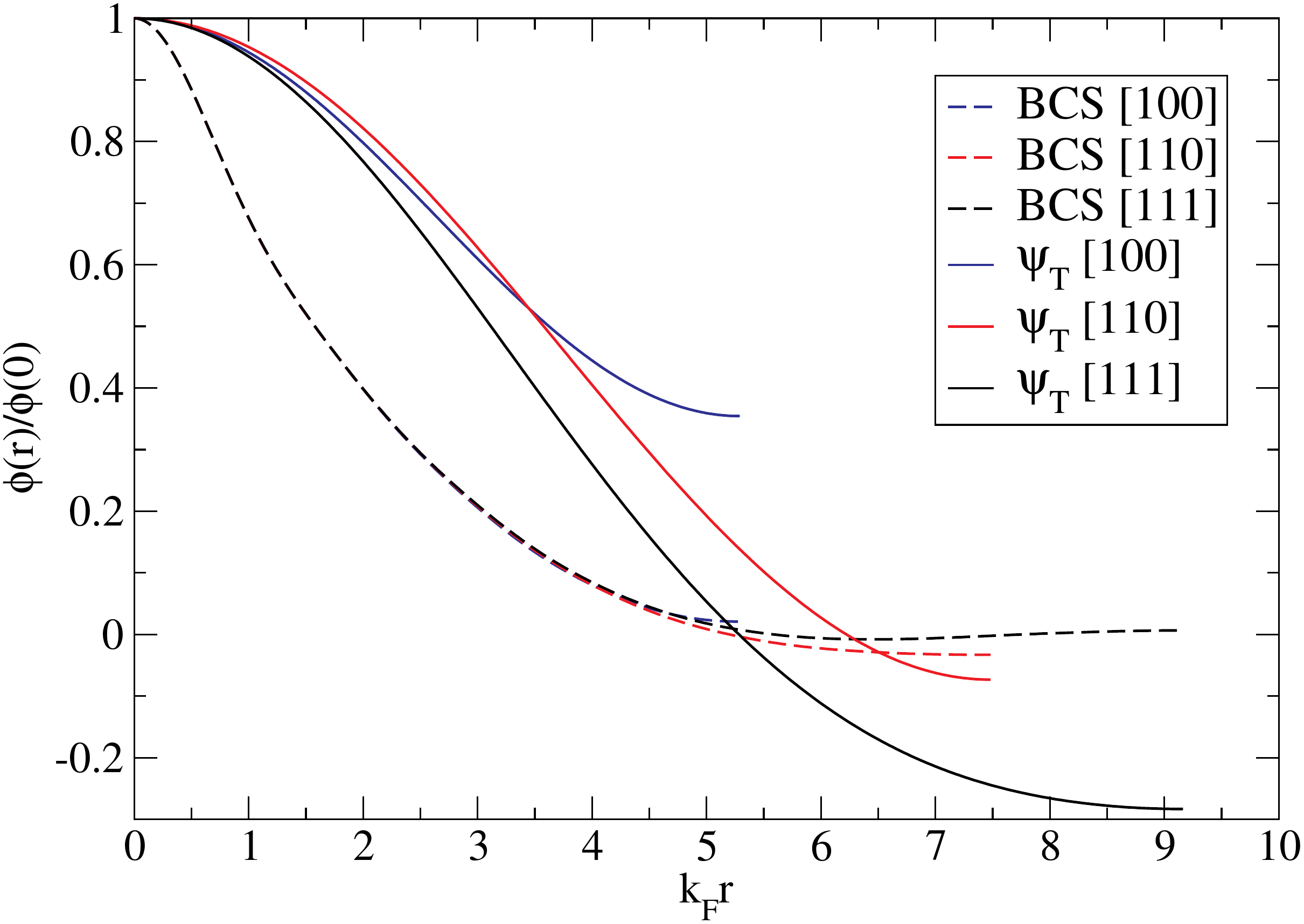}
\caption{The pairing function used in this work (solid lines) and the pairing function of the BCS ground state (dashed lines) in the $[100]$, $[110]$, and $[111]$ directions inside the box, for $N=40$ particles at  $k_\textrm{F}=0.4~\textrm{fm}^{-1}$.} \label{fig:pair-func}
\end{figure}

\section{Equation of State}\label{sec:results-energy}

We first examine the EoS of low-density NM. In Figure~\ref{fig:eos}, we present our AFDMC calculations with the better optimized trial wavefunction and we compare with the old AFDMC calculations from Refs.~\cite{Gandolfi:2008,Gandolfi:2009}, where correlated basis function calculations prescribed the pairing functions, and the results from Ref.~\cite{Gezerlis:2008}, where the DMC method was used instead and the pairing functions were determined by minimizing the trial state's energy. The results display the correct behavior at low densities, i.e.,~\mbox{$E/E_{\rm FG}\to 1$}, where pairing becomes less prevalent. 
Evidently, the better optimization of the pairing functions in the trial wave function of AFDMC provides a better description of superfluid NM yielding a EoS consistent with the DMC calculations of Ref.~\cite{Gezerlis:2008} for low and intermediate densities. Our work then extends these results to higher densities. 

\begin{figure}[H]
\includegraphics[width=.95\columnwidth]{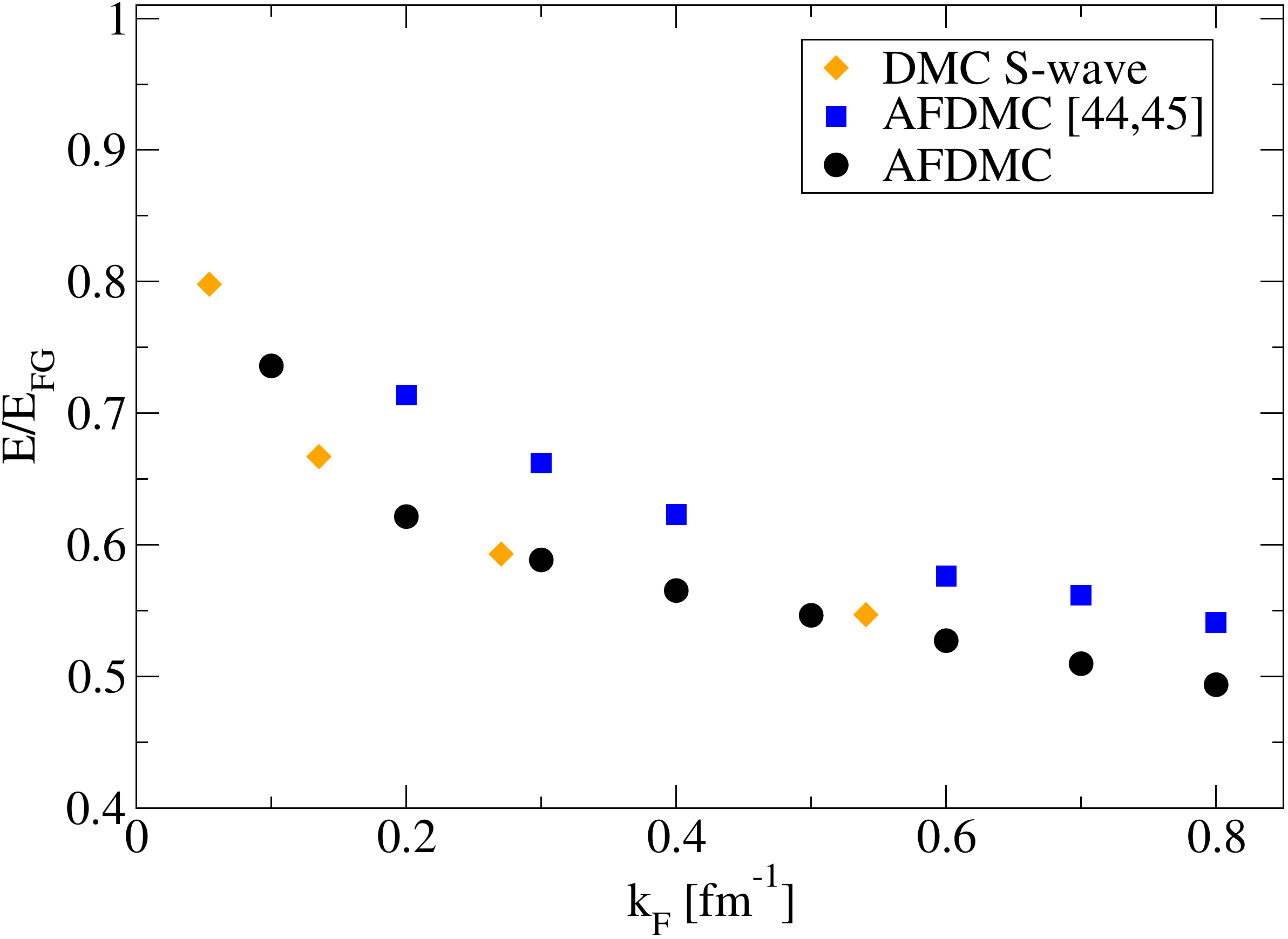}
\caption{The equation of state (EoS) of NM, in units of the free Fermi gas energy, as
a function of $k_\textrm{F}$. The results of this work (black circles) are compared to the old EOS of Ref.~\cite{Gandolfi:2008,Gandolfi:2009} (blue squares) and to 
quantum Monte Carlo (QMC) calculations of Ref.~\cite{Gezerlis:2008} (orange diamonds). 
}
\label{fig:eos}
\end{figure}

Our energy results have been obtained by simulating a system using $N=40$ neutrons under periodic boundary conditions. 

By choosing this specific particle number, we avoid closed-shell configurations e.g., $N=38$ or $66$. This functions to test that the pfaffian wave function with a properly optimized pairing functions that can reproduce open-shell configurations. It should be noted that, as demonstrated in Figure~\ref{fig:v2k}, superfluid systems do not yield a well-defined Fermi surface, and so all mentions to closed- and open-shell configurations refer to free systems with the same particle number. Certain closed-shell configurations are seen to produce minimum superfluid FSE (see Figure~\ref{fig:kinetic}) making them a common choice for QMC calculations~\cite{Gezerlis:2008, Gandolfi:2008}. Studying the TL with open-shell configurations would otherwise require the use of twist-averaged boundary conditions, as has been demonstrated for normal-state NM~\cite{Riz:2020} and superfluid NM~\cite{Palkanoglou:2021}. For some density, we changed $N$ from 40 to 66, without finding any dependency to $N$. This is because for such small densities, the box size is large enough to avoid FSE due to the truncation of the interaction. This has also been verified for ultra-cold Fermi gases with strong interactions~\cite{Forbes:2011}. The choice of $N=40$ is additionally motivated by overall efficiency: The optimization of the variational parameters becomes much harder for large $N$. As seen in Figure~\ref{fig:kinetic}, the FSE of the kinetic energy are smeared, compared to that of the free Fermi gas, due to the pairing correlations, providing smoother FSE to the total energy as well, as seen in the left panel of Figure~\ref{fig:fse}. From these calculations, we expect that the energy of a system with $N=40$ neutrons is within $\sim$2.5\% from its TL value.

\section{Pairing Gap}\label{sec:results-gap}
We have also performed calculations of the pairing gap using the better optimized trial state. The pairing gap is calculated using Equation~(\ref{eq:oes}) for an odd number of neutrons~$N$:
\begin{align}
\Delta(N) = E(N)-\frac{1}{2}\left[E(N+1) + E(N-1)\right] ~.
\label{eq:qmc.gap}
\end{align}

The results are shown in Figure~\ref{fig:gap}.
Since the pairing gap is calculated as differences between the total energy of the system, the result is that the uncertainty associated grows with $N$ making the use of large systems impractical. For most of the results presented here, we chose $N=47$ to get a reliable pairing gap and to minimize any possible bias due to closed shells. This is particularly important for lower densities where the pairing is suppressed. We tested the consistency of our results by performing simulations
at different numbers of neutrons up to $N=68$. Additionally, based on PBCS calculations of the pairing gap, we expect the pairing gap at $N=47$ to be close to the TL. In fact, the FSE, which for low densities and at small $N$, appear to be insensitive to density variations, suggest that the pairing gap of a system with $N=47$ neutrons is within $\sim 5\%$ from the TL (see the right panel of Figure~\ref{fig:fse}).

The results of the pairing gap are presented in Figure~\ref{fig:gap} where we 
also show the pairing gaps of Ref.~\cite{Gezerlis:2008} calculated using DMC and the old AFDMC results
of Refs.~\cite{Gandolfi:2008,Gandolfi:2009}. At low densities, the AFDMC calculations are in reasonable agreement with DMC calculations.
For higher densities, the AFDMC results depart from the BCS prediction as the mean-field approximation becomes less accurate.  In this work, we find a sizable pairing gap up to $k_F\sim 1.3$ fm$^{-1}$, 
in contrast to the previous AFDMC calculations that seem to drop around $k_F=0.8~\textrm{fm}^{-1}$ because of the much
simpler variatinal wave function employed.
This again is a consequence of the better optimization of the pairing functions that we adopted
in this work; we are capturing more of the essential correlations found in the ground state of superfluid NM. Overall, we find a moderate suppression of the pairing gap predicted by the mean-field BCS.

\begin{figure}[H]
\includegraphics[width=.95\columnwidth]{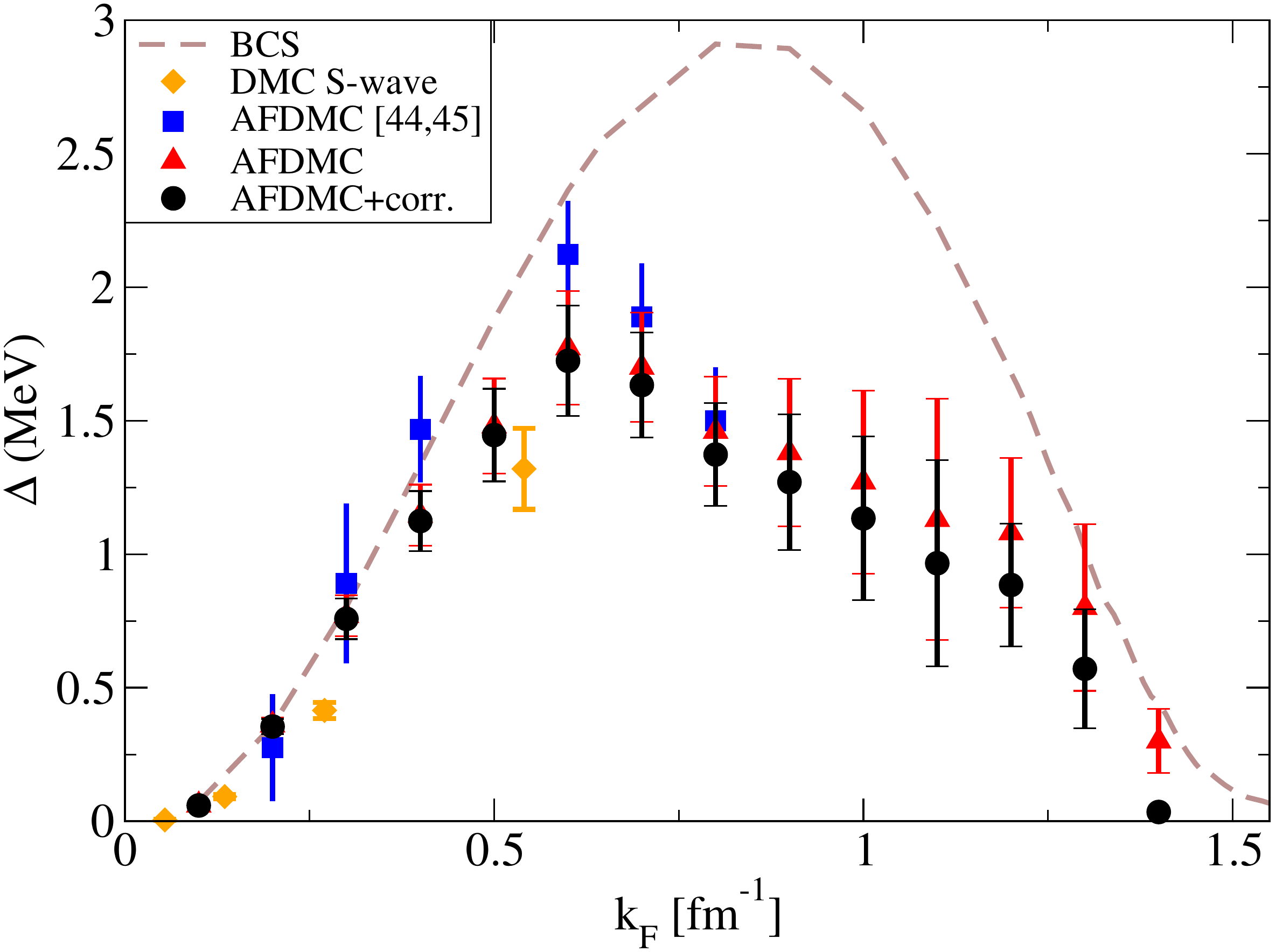}
\caption{The pairing gap calculated at constant density with (black circles) and without (red triangles) the correction described in the text. Our calculations are compared to the old AFDMC results of Refs.~\cite{Gandolfi:2008,Gandolfi:2009}, and to those of Ref.~\cite{Gezerlis:2008}. The brown dashed curve shows the pairing gap predicted by BCS.
}
\label{fig:gap}
\end{figure}

The calculations of the energies in Equation~(\ref{eq:qmc.gap}) are typically performed by simulating the system at a constant density in order to minimize
the FSE due to the truncation of the potential energy in the periodic box.
However, for the free Fermi gas, this procedure would give a non-zero pairing gap. We corrected our results of the pairing gap by subtracting that of the free Fermi gas obtained at constant densities. The correction is negligible at low densities, as can be seen in Figure~\ref{fig:gap}, and within error bars at large densities. In a few cases, we  also performed simulations at a constant volume instead of constant density and the results are very similar to corrected constant-density ones.

In summary, we performed a detailed ab initio study of the $S$ wave pairing in low-density NM found in the inner crust of cold NSs. We calculated the EoS and the pairing gap by means of AFDMC simulations for a range of densities, using a variationally optimized trial state benchmarked against previous calculations at low densities. A study of the FSE within a symmetry-restored mean-field treatment shows that the AFDMC energies and pairing gaps are within $\sim$2.5\% and $\sim$5\% from the TL, respectively.  Our AFDMC pairing gaps show a modest suppression with respect to the mean-field BCS values. These results can be used in calculations of the thermal properties of NSs and be indirectly tested in cold atom experiments utilizing the universality of the unitary Fermi gas. 

\authorcontributions{All authors have contributed to the investigation and to the writing of the article. All authors have read and agreed to the published version of the manuscript.}

\funding{The work of J.C. and S.G. has been supported by the NUclear Computational 
Low-Energy Initiative (NUCLEI) SciDAC project and by the 
U.S.~Department of Energy, Office of Science, Office of Nuclear Physics, under 
contracts DE-AC52-06NA25396.
The work of S.G. has also been supported by the DOE Early Career research program. The work of K.E.S. has been supported by the NSF Award DBI-1565180.
The work of G.P. and A.G. has been supported by the Natural Sciences and Engineering Research Council (NSERC) of Canada,
the Canada Foundation for Innovation (CFI), and the
Early Researcher Award (ERA) program of the Ontario
Ministry of Research, Innovation, and Science. Computational resources were provided by SHARCNET and
NERSC.
This research used resources provided by
the Los Alamos National Laboratory Institutional Computing Program,
which is supported by the U.S. Department of Energy National Nuclear
Security Administration under contract no. 89233218CNA000001.}


\conflictsofinterest{The authors declare no conflict of interest.}

\end{paracol}
\reftitle{References}


\begin{thebibliography}{999}

\end{thebibliography}


\begin{thebibliography}{999}

\bibitem[Gandolfi \em{et~al.}(2015)Gandolfi, Gezerlis, and
  Carlson]{Gandolfi:2015}
Gandolfi, S.; Gezerlis, A.; Carlson, J.
\newblock Neutron Matter from Low to High Density.
\newblock {\em Annu. Rev. Nucl. Part. Sci.} {\bf 2015}, {\em 65},~303--328.

\bibitem[{Ketterle} and {Zwierlein}(2008)]{Ketterle:2008}
{Ketterle}, W.; {Zwierlein}, M.W.
\newblock {Making, probing and understanding ultracold Fermi gases}.
\newblock {\em Riv. Nuovo Cim.} {\bf 2008}, {\em 31},~247--422.

\bibitem[Strinati \em{et~al.}(2021)Strinati, Pieri, Röpke, Schuck, and
  Urban]{Ramanan:2021}
Strinati, G.C.; Pieri, P.; Röpke, G.; Schuck, P.; Urban, M.
\newblock Pairing in pure neutron matter.
\newblock {\em Eur. Phys. J. Spec. Top.} {\bf 2021}, {\em 230},~567--577.

\bibitem[Dean and Hjorth-Jensen(2003)]{Dean:2003}
Dean, D.J.; Hjorth-Jensen, M.
\newblock Pairing in nuclear systems: from neutron stars to finite nuclei.
\newblock {\em Rev. Mod. Phys.} {\bf 2003}, {\em 75},~607--656.

\bibitem[Page \em{et~al.}(2004)Page, Lattimer, Prakash, and Steiner]{Page:2004}
Page, D.; Lattimer, J.M.; Prakash, M.; Steiner, A.W.
\newblock Minimal Cooling of Neutron Stars: A New Paradigm.
\newblock {\em Astrophys. J. Suppl. Sci.} {\bf 2004}, {\em 155},~623--650.

\bibitem[Page \em{et~al.}(2011)Page, Prakash, Lattimer, and Steiner]{Page:2011}
Page, D.; Prakash, M.; Lattimer, J.M.; Steiner, A.W.
\newblock Rapid Cooling of the Neutron Star in Cassiopeia A Triggered by
  Neutron Superfluidity in Dense Matter.
\newblock {\em Phys. Rev. Lett.} {\bf 2011}, {\em 106},~081101.

\bibitem[Yakovlev and Pethick(2004)]{Yakovlev:2004}
Yakovlev, D.; Pethick, C.
\newblock Neutron Star Cooling.
\newblock {\em Annu. Rev. Astron. Astr.} {\bf 2004}, {\em 42},~169--210.

\bibitem[Watanabe and Pethick(2017)]{Watanabe:2017}
Watanabe, G.; Pethick, C.J.
\newblock Superfluid Density of Neutrons in the Inner Crust of Neutron Stars:
  New Life for Pulsar Glitch Models.
\newblock {\em Phys. Rev. Lett.} {\bf 2017}, {\em 119},~062701.

\bibitem[Lattimer and Prakash(2016)]{Lattimer:2016}
Lattimer, J.M.; Prakash, M.
\newblock The equation of state of hot, dense matter and neutron stars.
\newblock {\em Phys. Rep.} {\bf 2016}, {\em 621},~127--164.

\bibitem[Chamel(2017)]{Chamel:2017}
Chamel, N.
\newblock Entrainment in Superfluid Neutron-Star Crusts: Hydrodynamic
  Description and Microscopic Origin.
\newblock {\em J. Low Temp. Phys.} {\bf 2017}, {\em 189},~328--360.

\bibitem[Pilati and Giorgini(2008)]{Pilati:2008}
Pilati, S.; Giorgini, S.
\newblock Phase Separation in a Polarized Fermi Gas at Zero Temperature.
\newblock {\em Phys. Rev. Lett.} {\bf 2008}, {\em 100},~030401.

\bibitem[Strinati \em{et~al.}(2018)Strinati, Pieri, Röpke, Schuck, and
  Urban]{Strinati:2018}
Strinati, G.C.; Pieri, P.; Röpke, G.; Schuck, P.; Urban, M.
\newblock The BCS–BEC crossover: From ultra-cold Fermi gases to nuclear
  systems.
\newblock {\em Phys. Rep.} {\bf 2018}, {\em 738},~1--76.

\bibitem[Forbes \em{et~al.}(2011)Forbes, Gandolfi, and Gezerlis]{Forbes:2011}
Forbes, M.M.; Gandolfi, S.; Gezerlis, A.
\newblock Resonantly Interacting Fermions in a Box.
\newblock {\em Phys. Rev. Lett.} {\bf 2011}, {\em 106},~235303.

\bibitem[Galea \em{et~al.}(2016)Galea, Dawkins, Gandolfi, and
  Gezerlis]{Galea:2016}
Galea, A.; Dawkins, H.; Gandolfi, S.; Gezerlis, A.
\newblock Diffusion Monte Carlo study of strongly interacting two-dimensional
  Fermi gases.
\newblock {\em Phys. Rev. A} {\bf 2016}, {\em 93},~023602.

\bibitem[Carlson \em{et~al.}(2011)Carlson, Gandolfi, Schmidt, and
  Zhang]{Carlson:2011}
Carlson, J.; Gandolfi, S.; Schmidt, K.E.; Zhang, S.
\newblock Auxiliary-field quantum Monte Carlo method for strongly paired
  fermions.
\newblock {\em Phys. Rev. A} {\bf 2011}, {\em 84},~061602.

\bibitem[Magierski \em{et~al.}(2009)Magierski, Wlaz\l{}owski, Bulgac, and
  Drut]{Magierski:2009}
Magierski, P.; Wlaz\l{}owski, G.; Bulgac, A.; Drut, J.E.
\newblock Finite-Temperature Pairing Gap of a Unitary Fermi Gas by Quantum
  Monte Carlo Calculations.
\newblock {\em Phys. Rev. Lett.} {\bf 2009}, {\em 103},~210403.

\bibitem[Zielinski \em{et~al.}(2020)Zielinski, Ross, and
  Gezerlis]{Zielinski:2020}
Zielinski, T.; Ross, B.; Gezerlis, A.
\newblock Pairing in two-dimensional Fermi gases with a coordinate-space
  potential.
\newblock {\em Phys. Rev. A} {\bf 2020}, {\em 101},~033601.

\bibitem[Tajima \em{et~al.}(2017)Tajima, van Wyk, Hanai, Kagamihara, Inotani,
  Horikoshi, and Ohashi]{Tajima:2017}
Tajima, H.; van Wyk, P.; Hanai, R.; Kagamihara, D.; Inotani, D.; Horikoshi, M.;
  Ohashi, Y.
\newblock Strong-coupling corrections to ground-state properties of a
  superfluid Fermi gas.
\newblock {\em Phys. Rev. A} {\bf 2017}, {\em 95},~043625.

\bibitem[Bartenstein \em{et~al.}(2005)Bartenstein, Altmeyer, Riedl, Geursen,
  Jochim, Chin, Denschlag, Grimm, Simoni, Tiesinga, Williams, and
  Julienne]{Bartenstein:2005}
Bartenstein, M.; Altmeyer, A.; Riedl, S.; Geursen, R.; Jochim, S.; Chin, C.;
  Denschlag, J.H.; Grimm, R.; Simoni, A.; Tiesinga, E.;  et~al.
\newblock Precise Determination of $^{6}\mathrm{Li}$ Cold Collision Parameters
  by Radio-Frequency Spectroscopy on Weakly Bound Molecules.
\newblock {\em Phys. Rev. Lett.} {\bf 2005}, {\em 94},~103201.

\bibitem[Ho(2004)]{Ho:2004}
Ho, T.L.
\newblock Universal Thermodynamics of Degenerate Quantum Gases in the Unitarity
  Limit.
\newblock {\em Phys. Rev. Lett.} {\bf 2004}, {\em 92},~090402.

\bibitem[Ohashi \em{et~al.}(2020)Ohashi, Tajima, and {van Wyk}]{Ohashi:2020}
Ohashi, Y.; Tajima, H.; {van Wyk}, P.
\newblock BCS–BEC crossover in cold atomic and in nuclear systems.
\newblock {\em Prog. Part. Nucl. Phys.} {\bf 2020}, {\em
  111},~103739.

\bibitem[Ding \em{et~al.}(2016)Ding, Rios, Dussan, Dickhoff, Witte, Carbone,
  and Polls]{Ding:2016}
Ding, D.; Rios, A.; Dussan, H.; Dickhoff, W.H.; Witte, S.J.; Carbone, A.;
  Polls, A.
\newblock Pairing in high-density neutron matter including short- and
  long-range correlations.
\newblock {\em Phys. Rev. C} {\bf 2016}, {\em 94},~025802.

\bibitem[Pavlou \em{et~al.}(2017)Pavlou, Mavrommatis, Moustakidis, and
  Clark]{Pavlou:2017}
Pavlou, G.; Mavrommatis, E.; Moustakidis, C.; Clark, J.
\newblock Microscopic study of 1 S 0 superfluidity in dilute neutron matter.
\newblock {\em Eur. Phys. J. A} {\bf 2017}, {\em 53},~1--9.

\bibitem[Krotscheck and Clark(1980)]{Krotscheck:1980}
Krotscheck, E.; Clark, J.
\newblock Studies in the method of correlated basis functions:(III). Pair
  condensation in strongly interacting Fermi systems.
\newblock {\em Nucl. Phys. A} {\bf 1980}, {\em 333},~77--115.

\bibitem[Fan and Krotscheck(2018)]{Fan:2018}
Fan, H.H.; Krotscheck, E.
\newblock An analysis of variational wave function for the pairing problem in
  strongly correlated system.
\newblock  \emph{J.Phys. Conf. Ser.}  \textbf{2018}, \emph{1041}, 012010.

\bibitem[Cao \em{et~al.}(2006)Cao, Lombardo, and Schuck]{Cao:2006}
Cao, L.G.; Lombardo, U.; Schuck, P.
\newblock Screening effects in superfluid nuclear and neutron matter within
  Brueckner theory.
\newblock {\em Phys. Rev. C} {\bf 2006}, {\em 74},~064301.

\bibitem[Wambach \em{et~al.}(1993)Wambach, Ainsworth, and Pines]{Wambach:1993}
Wambach, J.; Ainsworth, T.; Pines, D.
\newblock Quasiparticle interactions in neutron matter for applications in
  neutron stars.
\newblock {\em Nucl. Phys. A} {\bf 1993}, {\em 555},~128--150.

\bibitem[Rickayzen(1965)]{Rickayzen:book}
Rickayzen, G.
\newblock {\em Theory of Superconductivity}; Wiley, 1965.

\bibitem[Palkanoglou \em{et~al.}(2020)Palkanoglou, Diakonos, and
  Gezerlis]{Palkanoglou:2020}
Palkanoglou, G.; Diakonos, F.K.; Gezerlis, A.
\newblock From odd-even staggering to the pairing gap in neutron matter.
\newblock {\em Phys. Rev. C} {\bf 2020}, {\em 102},~064324.

\bibitem[Palkanoglou and Gezerlis(2021)]{Palkanoglou:2021}
Palkanoglou, G.; Gezerlis, A.
\newblock Superfluid Neutron Matter with a Twist.
\newblock {\em Universe} {\bf 2021}, {\em 7}, {\em 2}.

\bibitem[Gezerlis and Carlson(2010)]{Gezerlis:2010}
Gezerlis, A.; Carlson, J.
\newblock Low-density neutron matter.
\newblock {\em Phys. Rev. C} {\bf 2010}, {\em 81},~025803.

\bibitem[Dietrich \em{et~al.}(1964)Dietrich, Mang, and Pradal]{Dietrich:1964}
Dietrich, K.; Mang, H.J.; Pradal, J.H.
\newblock Conservation of Particle Number in the Nuclear Pairing Model.
\newblock {\em Phys. Rev.} {\bf 1964}, {\em 135},~B22--B34.

\bibitem[Bayman(1960)]{Bayman:1960}
Bayman, B.
\newblock A derivation of the pairing-correlation method.
\newblock {\em Nucl. Phys.} {\bf 1960}, {\em 15},~33--38.

\bibitem[Wiringa and Pieper(2002)]{Wiringa:2002}
Wiringa, R.B.; Pieper, S.C.
\newblock Evolution of Nuclear Spectra with Nuclear Forces.
\newblock {\em Phys. Rev. Lett.} {\bf 2002}, {\em 89},~182501.

\bibitem[Pudliner \em{et~al.}(1995)Pudliner, Pandharipande, Carlson, and
  Wiringa]{Pudliner:1995}
Pudliner, B.S.; Pandharipande, V.R.; Carlson, J.; Wiringa, R.B.
\newblock Quantum {M}onte {C}arlo Calculations of ${A} \le{}6$ Nuclei.
\newblock {\em Phys. Rev. Lett.} {\bf 1995}, {\em 74},~4396--4399.

\bibitem[Fujita and Miyazawa(1957)]{Fujita:1957}
Fujita, J.i.; Miyazawa, H.
\newblock {Pion Theory of Three-Body Forces}.
\newblock {\em Prog. Theor. Phys.} {\bf 1957}, {\em 17},~360--365.

\bibitem[Gezerlis \em{et~al.}(2014)Gezerlis, Tews, Epelbaum, Freunek, Gandolfi,
  Hebeler, Nogga, and Schwenk]{Gezerlis:2014}
Gezerlis, A.; Tews, I.; Epelbaum, E.; Freunek, M.; Gandolfi, S.; Hebeler, K.;
  Nogga, A.; Schwenk, A.
\newblock {Local chiral effective field theory interactions and quantum Monte
  Carlo applications}.
\newblock {\em Phys. Rev. C} {\bf 2014}, {\em 90},~054323.

\bibitem[Lynn \em{et~al.}(2016)Lynn, Tews, Carlson, Gandolfi, Gezerlis,
  Schmidt, and Schwenk]{Lynn:2016}
Lynn, J.E.; Tews, I.; Carlson, J.; Gandolfi, S.; Gezerlis, A.; Schmidt, K.E.;
  Schwenk, A.
\newblock {Chiral Three-Nucleon Interactions in Light Nuclei, Neutron-$\alpha$
  Scattering, and Neutron Matter}.
\newblock {\em Phys. Rev. Lett.} {\bf 2016}, {\em 116},~062501.

\bibitem[Gezerlis and Carlson(2008)]{Gezerlis:2008}
Gezerlis, A.; Carlson, J.
\newblock Strongly paired fermions: Cold atoms and neutron matter.
\newblock {\em Phys. Rev. C} {\bf 2008}, {\em 77},~032801.

\bibitem[Foulkes \em{et~al.}(2001)Foulkes, Mitas, Needs, and
  Rajagopal]{Foulkes:2001}
Foulkes, W.M.C.; Mitas, L.; Needs, R.J.; Rajagopal, G.
\newblock Quantum Monte Carlo simulations of solids.
\newblock {\em Rev. Mod. Phys.} {\bf 2001}, {\em 73},~33--83.

\bibitem[Carlson \em{et~al.}(2015)Carlson, Gandolfi, Pederiva, Pieper,
  Schiavilla, Schmidt, and Wiringa]{Carlson:2015}
Carlson, J.; Gandolfi, S.; Pederiva, F.; Pieper, S.C.; Schiavilla, R.; Schmidt,
  K.E.; Wiringa, R.B.
\newblock {Quantum Monte Carlo methods for nuclear physics}.
\newblock {\em Rev. Mod. Phys.} {\bf 2015}, {\em 87},~1067--1118.

\bibitem[Lonardoni \em{et~al.}(2018)Lonardoni, Gandolfi, Lynn, Petrie, Carlson,
  Schmidt, and Schwenk]{Lonardoni:2018}
Lonardoni, D.; Gandolfi, S.; Lynn, J.E.; Petrie, C.; Carlson, J.; Schmidt,
  K.E.; Schwenk, A.
\newblock {Auxiliary field diffusion Monte Carlo calculations of light and
  medium-mass nuclei with local chiral interactions}.
\newblock {\em Phys. Rev. C} {\bf 2018}, {\em 97},~044318.

\bibitem[Fabrocini \em{et~al.}(2005)Fabrocini, Fantoni, Illarionov, and
  Schmidt]{Fabrocini:2005}
Fabrocini, A.; Fantoni, S.; Illarionov, A.Y.; Schmidt, K.E.
\newblock $^1{S}_0$ Superfluid Phase Transition in Neutron Matter with
  Realistic Nuclear Potentials and Modern Many-Body Theories.
\newblock {\em Phys. Rev. Lett.} {\bf 2005}, {\em 95},~192501.

\bibitem[Gandolfi \em{et~al.}(2008)Gandolfi, Illarionov, Fantoni, Pederiva, and
  Schmidt]{Gandolfi:2008}
Gandolfi, S.; Illarionov, A.Y.; Fantoni, S.; Pederiva, F.; Schmidt, K.E.
\newblock Equation of State of Superfluid Neutron Matter and the Calculation of
  the $^{1}S_{0}$ Pairing Gap.
\newblock {\em Phys. Rev. Lett.} {\bf 2008}, {\em 101},~132501.

\bibitem[Gandolfi \em{et~al.}(2009)Gandolfi, Illarionov, Pederiva, Schmidt, and
  Fantoni]{Gandolfi:2009}
Gandolfi, S.; Illarionov, A.Y.; Pederiva, F.; Schmidt, K.E.; Fantoni, S.
\newblock Equation of state of low-density neutron matter, and the
  ${}^{1}{S}_{0}$ pairing gap.
\newblock {\em Phys. Rev. C} {\bf 2009}, {\em 80},~045802.

\bibitem[Carlson \em{et~al.}(2003)Carlson, Chang, Pandharipande, and
  Schmidt]{Carlson:2003}
Carlson, J.; Chang, S.Y.; Pandharipande, V.R.; Schmidt, K.E.
\newblock Superfluid Fermi Gases with Large Scattering Length.
\newblock {\em Phys. Rev. Lett.} {\bf 2003}, {\em 91},~050401.

\bibitem[Fabrocini \em{et~al.}(2008)Fabrocini, Fantoni, Illarionov, and
  Schmidt]{Fabrocini:2008}
Fabrocini, A.; Fantoni, S.; Illarionov, A.Y.; Schmidt, K.E.
\newblock S--pairing in neutron matter. I.Correlated Basis Function Theory.
\newblock {\em Nucl. Phys. A} {\bf 2008}, {\em 803},~137.

\bibitem[Sorella(2001)]{Sorella:2001}
Sorella, S.
\newblock {Generalized Lanczos algorithm for variational quantum Monte Carlo}.
\newblock {\em Phys. Rev. B} {\bf 2001}, {\em 64},~024512.

\bibitem[Riz \em{et~al.}(2020)Riz, Pederiva, Lonardoni, and Gandolfi]{Riz:2020}
Riz, L.; Pederiva, F.; Lonardoni, D.; Gandolfi, S.
\newblock Spin Susceptibility in Neutron Matter from Quantum Monte Carlo
  Calculations.
\newblock {\em Particles} {\bf 2020}, {\em 3},~706--718.

\end{thebibliography}
\end{document}